\documentclass[a4paper,11pt]{article}
\usepackage{jinstpub}
\usepackage{lineno}

\usepackage{amssymb}
\usepackage{graphicx}
\usepackage{subcaption}
\usepackage{amsmath} 
\usepackage{enumerate} 
\usepackage{booktabs}
\usepackage{tabularx}
\usepackage{array}
\usepackage{rotating} 
\usepackage[utf8]{inputenc}

\title{\boldmath Design and Verification of the JUNO Liquid Filling Control System}

\author[a,1]{Jiajun Li,\note{Corresponding author.}}
\author[b]{Yuekun Heng,}
\author[a]{Tao Huang,}
\author[a,2]{Jiajie Ling,\note{Corresponding author.}}
\author[b]{Xiao Tang,}
\author[b,3]{Zhi Wu,\note{Corresponding author.}}
\author[a]{Chengfeng Yang,}
\author[c]{Fan Ye,}
\author[a]{Shiqi Zhang,}
\author[b]{Yinhong Zhang,}

\affiliation[a]{Sun Yat-sen University, No. 135, Xingang Xi Road, Guangzhou, China}
\affiliation[b]{Institute of High Energy Physics, 19B Yuquan Road, China}
\affiliation[c]{Spallation Neutron Source Science Center, 1 Zhongziyuan Road, Dongguan ,China}

\emailAdd{lijj295@mail2.sysu.edu.cn (Jiajun Li)}
\emailAdd{lingjj5@mail.sysu.edu.cn (Jiajie Ling)}
\emailAdd{wuz@ihep.ac.cn.cn (Zhi Wu)}

\abstract{Jiangmen Underground Neutrino Observatory (JUNO) is a large-scale neutrino experiment with multiple physics goals including neutrino mass hierarchy, accurate measurement of neutrino oscillation parameters, neutrino detection from supernova, sun, and earth, etc. This paper presents the design, implementation, and verification of a high-reliability automated control system for the liquid Filling, Overflow, and Circulation system in the JUNO experiment. The system is built upon a Programmable Logic Controller architecture, integrated with high-precision sensors and actuators. It implements advanced control strategies, including Proportional-Integral-Derivative regulation, sequential logic, and safety interlocks, to achieve closed-loop control of critical parameters such as flow rate, liquid level, and pressure. Commissioning tests with both pure water and liquid scintillator demonstrate the system's exceptional performance, achieving flow control stability within 0.5\% of the setpoint with a rapid stabilization time. The robust design, featuring hardware redundancy and software safeguards, ensures the system meets the stringent requirements for the safe filling and long-term stable operation of JUNO's 20-kiloton central detector and provides a scalable reference for large underground fluid control experiments.}

\keywords{JUNO, central detector, liquid scintillator, FOC system, automatic control}

\begin{document}
\maketitle
\flushbottom

\section{Introduction}

The Jiangmen Underground Neutrino Observatory (JUNO)~\cite{JUNOphy_det} is the next generation neutrino experiment, located about 53 km from the Taishan and Yangjiang nuclear power plants in south China to optimize the precise measurement of the reactor neutrino oscillation parameters.
Its primary goal is to determine the neutrino mass ordering - a fundamental question with implications for neutrino mass origins and physics beyond the Standard Model - while also measuring oscillation parameters ($\theta_{12}$, $\Delta m_{21}^2$, $|\Delta m_{32}^2|$) with per-mil precision. JUNO can also advance the research associated with solar, atmospheric, supernova neutrinos, geoneutrinos, proton decay, etc. 

The JUNO Central Detector (CD) is the largest liquid scintillator (LS) detector with 20-kiloton LS contained by a 35.4 meter diameter acrylic sphere, achieving 3\% energy resolution at 1 MeV. The acrylic sphere and photo-multiplier tubes (PMT) are supported by a 41.1 meter in diameter stainless steel shell \cite{CD_paper}. The whole detector is submerged in 40-kiloton pure water (PW) in water pool (WP) for radiation shielding and active cosmic-ray muon tagging. 

Since the pioneering experiment of Reines and Cowan \cite{Cowan}, LS has been widely used in neutrino experiments \cite{LSreview} due to its homogeneity, ease of handling, cost-effectiveness, large-scale availability, and purifiability. JUNO's LS requires ultra-high purity standards \cite{mixing}: optical attenuation length $>20$ m, dust content $<10$ mg, and system leakage rate $\leq 10^{-6}$ mbar$\cdot$L$\cdot$s$^{-1}$ to minimize radioactive impurities ($^{238}\text{U}$, $^{232}\text{Th}$, $^{40}\text{K}$) and gaseous contaminants ($^{222}\text{Rn}$, $^{85}\text{Kr}$, $^{39}\text{Ar}$). To maximize scintillation light generation and propagation, the optimal LS composites linear alkyl benzene (LAB) with \(2.5 \, \text{g/L}\) diphenyloxazole (PPO) and \(3 \, \text{mg/L}\) bis(2-methylstyryl)benzene (bis-MSB) \cite{LSrece}. The JUNO LS production and purification system is developed for storing, filtering and distilling LAB, mixing PPO and bis-MSB with LAB, removing water-soluble and gaseous elements. It can provide purified LS at a rate of \(7\ \text{m}^3/\text{h}\) to fill into CD.

The filling, overflow, and circulation (FOC) system is in responsible for LS filling, overflowing or refilling, and online circulation for LS re-purification \cite{CD_paper}. Filling the CD with purified LS is the final step in the detector construction. The process must preserve the structural integrity of the detector and the chemical and optical properties of the LS. During detector filling and data taking, the system should also be able to maintain LS pressure and control the LS level for detector safety. To meet the above requirements, the FOC system is designed with an operation lifetime of 20 years.

\begin{figure}[ht]
 \centering
 \makebox[\textwidth][c]{\includegraphics[width=1\textwidth]{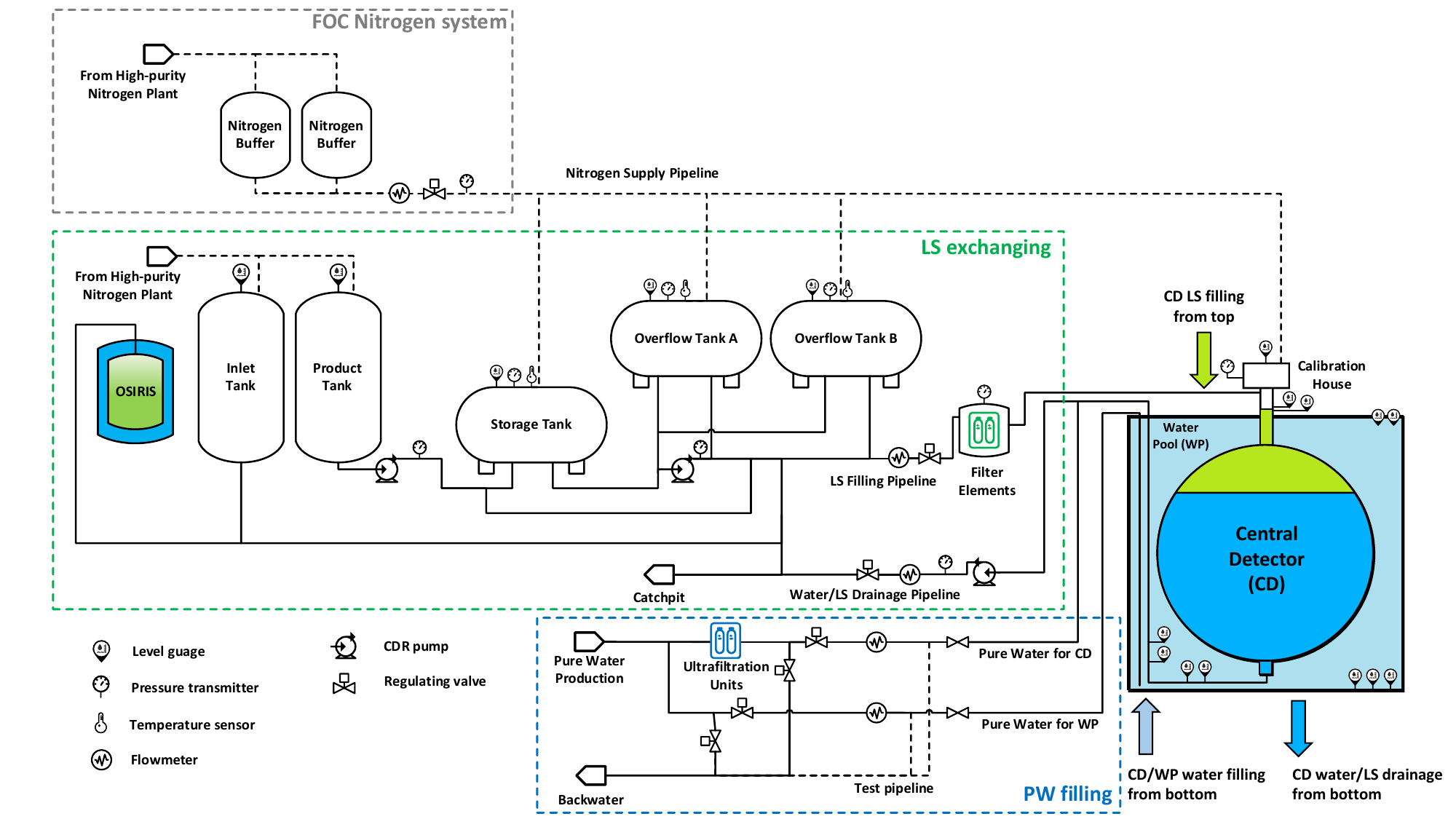}}%
 \caption{Primary P\&ID of the FOC system}
 \label{fig:foc_pid}
\end{figure}

The FOC system consists primarily of one storage tank as a buffer container for LS filling and circulation, and two overflow tanks compensating for LS volume fluctuations caused by temperature changes during detector operation. The system also includes pipes, top and bottom chimneys, nitrogen gas system preventing the LS from exposure to oxygen, moisture, and radon. It also features a control system to implement the above functions of FOC.
The corresponding Piping and Instrument Diagram (P\&ID) of the FOC system is shown in Figure \ref{fig:foc_pid}. 

This paper focuses on the control system of the FOC system. The paper is organized as follows. In Section 2, we describe the design, specifications, and principles related to the control system. In Section 3, we describe the principles of control logic and their application scenarios. We present the test performances of the control system in Section 4.

\section{System requirement and design}

\subsection{System requirement}

The filling process consists of: (1) synchronous pure water injection into both the CD and WP, (2) replacing pure water in the CD with LS. During the data-taking period, the system must maintain stable CD pressure while limiting liquid level fluctuations to within 20 cm (overflow and refilling condition), and support online LS circulation. To ensure these functions, the control system must meet several major requirements: precision, robustness, scalability, stability, anti-interference, safety and long time data traceability. 

Firstly, the control system of FOC must exhibit extremely high accuracy. The measurement deviations of all sensors shall be strictly constrained within the allowed range permitted by the process to ensure that every parameter during the filling and detector running accurately reflects the system's status. High-precision sensors provide a solid foundation for precise system control, with specific accuracy requirements such as the liquid level sensor measurement accuracy $\leq 0.2\%$ full scale (FS) and the flow sensor accuracy reaching $\pm 0.1\%$ FS. The system is designed to maintain deviations within $1\%$ between the setpoint and feedback value in a closed-loop control.
Pumps shall maintain rotational speed stability within $\pm 0.2\%$ of setpoint with $\leq 10$ seconds standby activation. Regulating valves require $\pm 1\%$ positioning accuracy, while on-off valves must achieve < 1 second response time and $< 10^{-6}\, \text{mbar}\cdot\text{L}\cdot\text{s}^{-1}$ leak rate.

Secondly, the control system must possess security and scalability. Specifically, it requires safety interlock mechanisms to mitigate risks to detectors during abnormal conditions. A physical emergency stop function is needed to ensure process termination even during system failures. For safety, we also use a "dual-confirmation" protocol to prevent hazardous operations caused by mistakes. Considering potential changes and upgrades in FOC system, the control system must feature modular expandability to avoid system redesigns.

In the end, the control system requires high stability, robustness and anti-interference capability. It must maintain precise control during prolonged continuous operation to avoid risks caused by control deviations and equipment failures. The system should be able to handle environmental fluctuations (e.g., electromagnetic interference, temperature variations, mechanical vibrations) or internal signal noise.

\subsection{System design}

\begin{figure}[ht]
 \centering
 \makebox[\textwidth][c]{\includegraphics[width=1\textwidth]{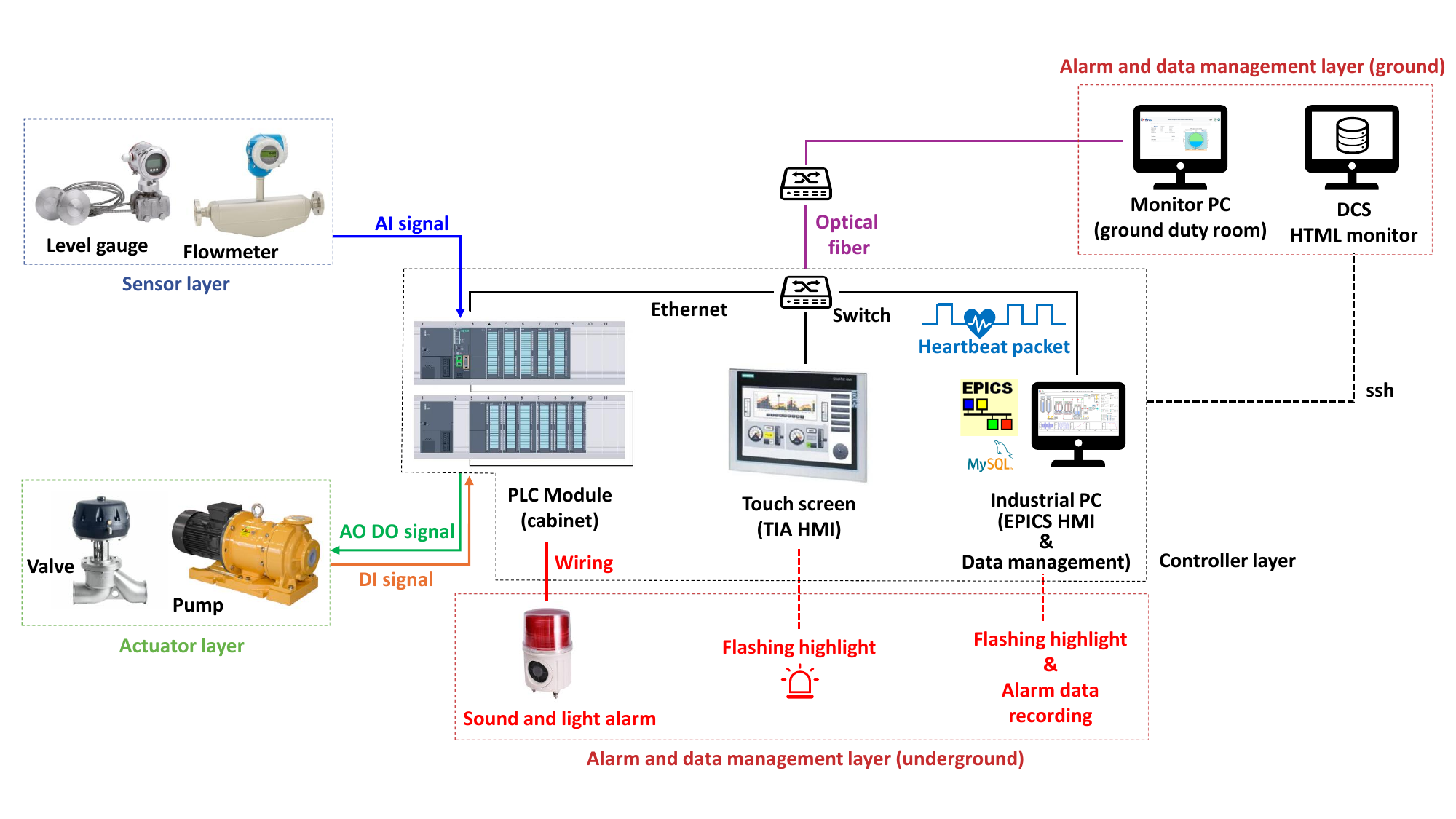}}%
 \caption{Layout of the control system}
 \label{fig:Automatic_control}
\end{figure}

Based on the above requirements, 
the layout of the control system is shown in Figure \ref{fig:Automatic_control}. The system integrates 35 sensors, 20 control valves, and 8 pump units, forming a sophisticated control network. In this configuration, the system manages more than 80 sets of analog input/output (I/O) control quantities and more than 150 digital I/O control and alarm signals. Therefore, it mainly consists of 4 layers: the sensor layer, the controller layer, the actuator layer, as well as the alarm and data storage layer.

\subsubsection{Sensor layer}

As the critical monitoring terminal, all sensors selected for the FOC system (including level, pressure, temperature, flow rate, etc.) are sourced from Endress+Hauser (E+H) \cite{endress_website}, renowned for their stability and reliability with an accuracy better than 0.2\% FS. The instrument selection underwent multiple rounds of demonstration, aligning with the system pipeline or tanks design and meeting the corresponding filling index requirements. For instance, the flowmeters were specifically chosen to match the LS filling rate of \(7\ \text{m}^3/\text{h}\) with optimal measuring range and precision. These devices have undergone rigorous acceptance testing, including factory tests, offline tests, and online tests. All sensors come pre-equipped with functions such as preset alarm thresholds, alarm signal output, and fault diagnostics. The surface roughness of sensor interfaces in contact with LS meets design requirements and maintains sufficient cleanliness.

\begin{table}[htbp]
    \centering
    \caption{Main sensor selection for control system}
    \label{tab:sensor_info}
    \begin{tabularx}{\linewidth}{p{3.7cm} c c X}
        \toprule
        \textbf{Sensor type} & \textbf{Quantity} & \textbf{Accuracy (\%FS)} & \textbf{Purpose} \\
        \midrule
        Electronic remote differential pressure level gauge & 4 & 0.20 & Measure CD liquid level (vertical tube \& horizontal tube)\\
        Laser level gauge & 1 & 0.10 & Measure CD liquid level \\
        Differential pressure level gauge & 2 & 0.20 & Measure liquid level in the upper chimney \\
        Static pressure liquid level gauge & 5 & 0.20 & Measure WP liquid level (bottom \& surface) \\
        Differential pressure level gauge & 5 & 0.20 & Measure tank liquid level \\
        Radar wave level gauge & 3 & 0.10 &  Measure tank liquid level \\
        Temperature sensor & 3 & 0.20 & Measure LS temperature inside tanks \\
        Coriolis mass flowmeter & 2 & 0.10 & Measure flow of exchanging LS/water \\
        Pressure transmitter & 3 & 0.15 & Measure tank gas pressure \\
        Pressure transmitter & 3 & 0.15 & Measure pump discharge pressure \\
        Pressure transmitter & 1 & 0.20 & Measure pressure of calibration house \\
        Double-flange pressure transmitter & 1 & 0.10 & Measure pressure difference before and after the filter \\
        Vortex flowmeter & 2 & 0.20 & Measure water flow during water filling period \\
        \bottomrule
    \end{tabularx}
\end{table}

The sensor specifications are summarized in Table \ref{tab:sensor_info}, which aligns with the system's requirements. Specifically, redundant sensors are deployed at critical points as follows: (1) For CD level monitoring: 2 pairs of electronic remote differential pressure gauges, 1 laser gauge, and 2 chimney differential pressure gauges; (2) For WP level monitoring: 5 static pressure gauges installed at different heights; (3) For tank level monitoring: dual level sensors for each tank; (4) For nitrogen pressure monitoring: 3 sensors installed at 3 connected tanks; (5) For LS temperature monitoring: four temperature sensors distributed throughout the pipelines and tanks.

These redundant instruments monitor the same parameter simultaneously, with primary and backup devices clearly defined. When the primary instrument fails, the standby device automatically switches to active mode within seconds, leveraging real-time diagnostics and failover algorithms. This hot standby mechanism ensures continuous and accurate data collection, effectively reducing measurement errors and system risks caused by single-instrument failures.

The sensor layer incorporates comprehensive anti-interference measurement to ensure signal reliability. All analog signal transmissions employ single-end grounded shielded twisted-pair cables to minimize electromagnetic interference, while LC filter modules are connected in series with the sensor power supply to suppress power ripple. Critical analog signals undergo through barrier isolators to prevent ground loops. These hardware protection measures work in tandem with the software filtering strategies described in Section 3.1, collectively achieving the goal of signal interference reduction.

Besides, a rigorous multi-stage calibration and acceptance procedure was implemented for all sensors. First, each device was supplied with factory calibration certificates, and their documented accuracy and stability met the experiment's stringent requirements. Then, our acceptance process comprised two phases: an off-line acceptance without load, and an on-line acceptance after installation. During these tests, the calibration of all sensors was verified. For those that could be calibrated within the FOC system itself---such as tank level gauges, pressure transmitters, temperature sensors, and pump discharge pressure sensors---we employed cross-comparison methods. A typical example is using two independent level gauges to validate each other's readings. The success of these calibration checks is further demonstrated by the system's performance during integrated testing, as detailed in Section 4. For critical sensors that could not be fully calibrated on-site prior to the JUNO detector filling---specifically, the CD and WP level gauges---a qualitative validation was performed using a small-scale mock-up. This setup allowed for cross-checking among multiple sensor groups, verifying their consistent response and functionality, thereby providing confidence in their readiness for the final operation.

\subsubsection{Controller layer}

The Siemens S7-300 series Programmable Logic Controller (PLC) is chosen as the core component of the control system. Equipped with a high-performance CPU, it has powerful computing and processing capabilities. It can analyze and judge various input signals in micro-second and then quickly issue control commands including equipment start-stop operations and process parameter adjustments. Its modular design enables flexible configuration to meet diverse control requirements while ensuring system reliability through independent module replacement, maintaining uninterrupted operation.

The modular configuration is chosen for FOC PLC as follows: one CPU module for computation, one interface module (IM) for data exchange, six 8-channel analog input (AI) modules for sensor signals, two 8-channel analog output (AO) modules for control, and three digital modules (two 32-channel DI and one 32-channel DO) for switch/alarm handling. 



\begin{figure}[ht]
 \centering
 \makebox[\textwidth][c]{\includegraphics[width=0.7\textwidth]{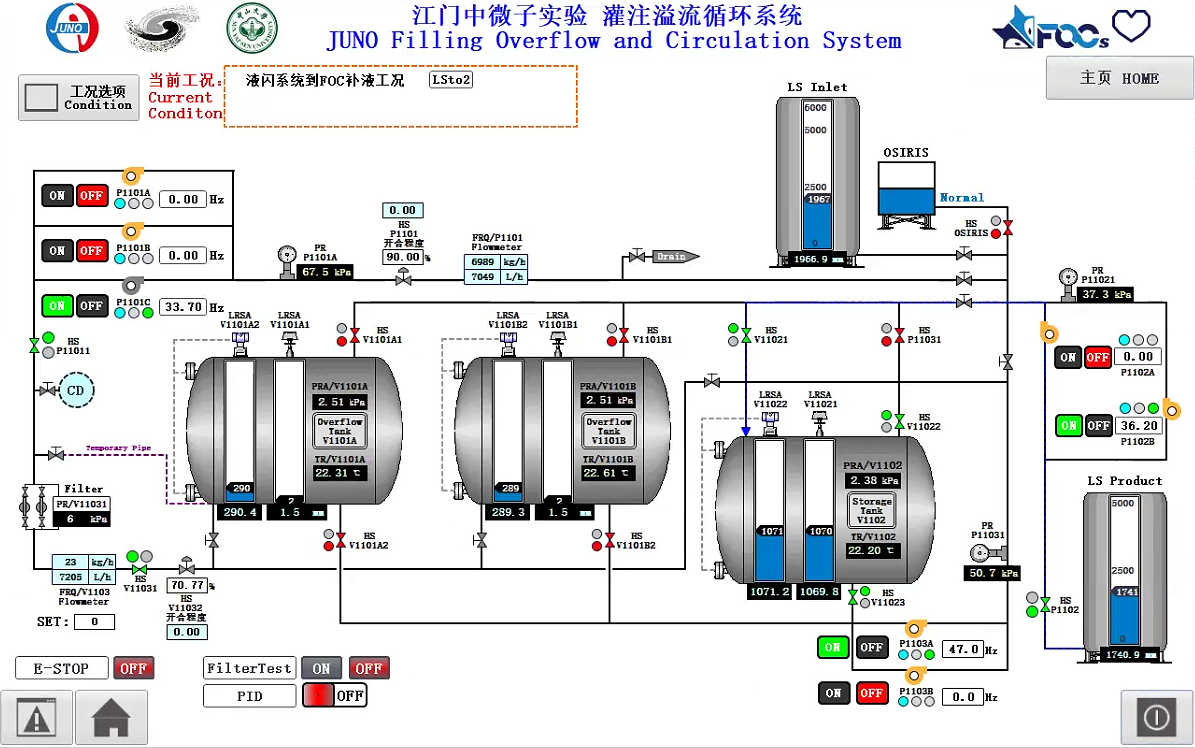}}%
 \caption{HMI based on TIA Portal}
 \label{fig:FOC HMI}
\end{figure}

The control software was developed using the Siemens Totally Integrated Automation (TIA) Portal \cite{TIA}, which provides integrated planning, diagnostic, and comprehensive monitoring capabilities. The primary human-machine interface (HMI) for local operation and real-time control is also developed and hosted within the TIA Portal environment, running on a dedicated industrial computer. This TIA-based HMI (shown in Figure \ref{fig:FOC HMI}) is the main interface for operators to perform control actions, such as adjusting valve openings and controlling pump frequency, and for displaying real-time process parameters during routine operations.

For integration into the broader JUNO Detector Control System (DCS), a separate monitoring and control program is developed based on the Experimental Physics and Industrial Control System (EPICS) \cite{EPICS} framework. The PLC, the TIA Portal HMI station, and the EPICS Input/Output Controller (IOC) -- running on a Linux-based industrial computer -- are interconnected via a dedicated industrial Ethernet switch.

The data synchronization between the PLC and the EPICS IOC is established through a network communication protocol (e.g., OPC UA or native Siemens S7), allowing the EPICS system to access all critical process variables. While the TIA HMI remains the primary interface for direct control, the EPICS channel serves two pivotal functions:
\begin{itemize}
    \item It enables comprehensive monitoring, historical data querying, and alarm management for the FOC system from the unified JUNO-wide DCS interface, facilitating coordinated operation with other subsystems.
    \item It implements a heartbeat monitoring program. The EPICS IOC continuously checks the operational status of the PLC. A loss of this heartbeat signal triggers a global alarm within the DCS, notifying operators of a potential PLC failure.
\end{itemize}

This dual-interface architecture ensures both robust, low-latency local control via the TIA ecosystem and seamless integration into the experiment's global monitoring and safety framework via EPICS.

\subsubsection{Actuator layer}

This layer mainly includes controllable on-off and regulating valves as well as pumps and frequency converter sets. SED valves \cite{SED}, CDR pumps \cite{CDR} and Yamei vacuum self-priming pumps \cite{Yamei} are used. The specification of pumps are listed in Table \ref{tab:pump_specs}. 
\begin{table}[htbp]
    \centering
    \caption{Pump specifications for control system} 
    \label{tab:pump_specs} 
    \begin{tabularx}{\linewidth}{p{4.2cm} c X} 
        \toprule
        \textbf{Pump type} & \textbf{Quantity} & \textbf{Purpose} \\
        \midrule
        Electromagnetic pump & 2 & Deliver LS from LS product tank to FOC system \\
        Electromagnetic pump & 2 & Deliver LS from FOC system to CD or return it to purification system \\
        Electromagnetic pump & 2 & Extract LS from CD and pump to the purification system for circulation \\
        Vacuum self-priming pump & 2 & Extract pure water from CD during LS exchanging \\
        \bottomrule
    \end{tabularx}
\end{table}
All electromagnetic pumps operate in the "one operation and one standby" configuration. Real-time pump status monitoring is achieved through digital feedback signals (ready, fault, running) and analog feedback of the actual operating frequency, which are compared against control signals. Pressure sensors downstream of pumps monitor hydraulic pressure to prevent overpressure, while upstream level interlocks safeguard against dry running. All valves come with ample spare parts and manual bypass valves. On-off valves provide “opened” or “closed” status feedback signals, while regulating valves offer analog feedback on opening degrees. Both pumps and valves have additional monitoring methods include local inverter displays and regular operator rounds, ensuring system stability. 

The actuators underwent rigorous acceptance testing to verify their performance aligns with the control system's requirements. Regulating valves were characterized to establish their flow coefficient curves, ensuring the PID control loops can accurately relate valve opening percentage to flow rate. The positioning feedback of these valves is continuously monitored, and any significant deviation from the commanded signal triggers a maintenance alert.

Similarly, the electromagnetic pumps were tested to confirm their rotational speed stability and response time under various load conditions. This characterization process for both valves and pumps is intrinsically coupled with the system performance tests described in Section 4. The stable flow control and precise pressure maintenance achieved during those tests (e.g., as summarized in Table~\ref{tab:performance_summary}) serve as the ultimate validation of the actuator calibration and integration. This proactive approach, combined with real-time status monitoring and the "one operation and one standby" configuration, minimizes the uncertainty in system actuation and provides a high-degree of operational reliability.

\subsubsection{Alarm and data management layer}

The control system integrates a variety of alarm methods. In the control cabinet, an industrial sound and light alarm siren horn is installed. Once the system malfunctions, the alarm device will immediately trigger audiovisual alerts on-site operators to deal with it in a timely manner. The HMI displays anomalies through red flashing widgets and pop-up alerts for rapid identification. The physical emergency stop button ensures immediate system shutdown under extreme conditions.

In terms of data management, all data related to FOC are properly stored. The system employs a dual data storage architecture, with local hard disk storage enabling real-time analysis and troubleshooting, while synchronized DCS storage facilitates distributed access and multi-department data sharing. This integrated approach, coupled with web-based remote monitoring capabilities, ensures comprehensive data traceability. 

\section{Control logic}

\subsection{Architecture of the control logic}

Building on the hardware architecture described in Section 2.2, drawing on the hierarchical control theory ISA-88 \cite{isa88}, the control system employs a closed-loop control framework through three functional layers: detection, control, and monitoring \& safety. This hierarchical design enables precise regulation of LS parameters while maintaining system stability during long-term operation.

\begin{itemize}

\item Detection layer: signal acquisition and processing

This layer interfaces with high-reliability sensors to collect real-time process parameters, employing a hybrid filtering strategy to mitigate noise. Arithmetic mean filtering reduces high-frequency fluctuations such as radar level gauge noise from 30 mm to 5 mm, while the time hysteresis filter combines dual conditions of threshold and duration, avoiding false triggers caused by transient interference signals of Coriolis flow meters.

\item Control layer: algorithm implementation

The control layer operates through four complementary control logic. Continuous control via Proportional-Integral-Derivative (PID) algorithms dynamically adjusts valve opening and pumping speed. State-machine-based sequential control enforces step-by-step execution, such as initiating LS exchanging only after tank level and temperature thresholds are met. The layer also integrates split-range control (which coordinates multiple actuators via a single controller) and selective control (which prioritizes critical parameters under competing constraints). Boolean logic interlocks provide additional protection, triggering actions such as shutting down pumps automatically if pressure goes beyond the safety limits. Detailed design and application scenarios will be introduced in Section 3.2.

\item Monitoring and safety layer: redundant safeguard

Detailed specifications of the monitoring and safety layer have been elaborated in Section 2.2. This layer incorporates HMI monitoring and multiple alarm modes, enabling operators to monitor the process parameters and manually intervene in the control logic to trigger critical actions timely during system failures.

\end{itemize}

\subsection{Typical control logic types and application scenarios}

Multiple typical control logic functions guarantee the operation of the FOC system safely and efficiently, which includes: PID control, sequential control, safety interlock, and split-range \& selective control logic.

\begin{itemize}
\item PID control
    
Following Ziegler-Nichols tuning rules \cite{pidcon}, the PID control logic is primarily used for the closed-loop control of continuous variables such as flow rate. In the LS filling process, precise control of the liquid flow rate is critical. An excessive flow rate may undermine the structural safety of the CD, while an extremely slow flow will reduce the filling efficiency. The logical principle of PID control is described as follows:
\[
\label{eq:1} 
u(t) = K_p \left( e(t) + \frac{1}{T_i} \int_{0}^{t} e(\tau)d\tau + T_d \frac{de(t)}{dt} \right), \tag{1}
\] 
where $u(t)$ denotes the control output signal to actuators (e.g., regulating valves), $K_p$ is the proportional gain, which is directly scaled to the current deviation $e(t)$ (setpoint minus actual value) for immediate response, $T_i$ (integral time constant) accumulates past deviations to eliminate steady-state errors, ensuring long-term accuracy, while $T_d$ (derivative time constant) predicts system trends by reacting to the rate of change of the deviation, damping overshoots. 

Taking LS filling control as an example: the PID controller dynamically adjusts the flow through three coordinated actions. When the actual flow falls below the setpoint, the proportional term immediately boosts the output. The derivative term counters rapid flow increases by generating opposing signals based on the rate of deviation change, preventing overshoot. Simultaneously, the integral term continuously eliminates residual errors - during initial filling, strong proportional action dominates, while the integral term handles fine adjustments as flow nears the target, with derivative action maintaining stability throughout the transition.

\item Sequential control
 
The sequential control logic employs a state-machine design to enforce strict stepwise execution of operations through condition-action chains, where each process phase has defined prerequisites and actions. For instance, LS exchanging initiation requires the storage tank to reach a preset level while meeting both the LS temperature specifications and calibration house pressure threshold, with the system only ramping up exchange speed after verifying stable low-flow filling to ensure detector safety.

\item Safety interlock

The safety interlock logic mitigates accidents by sensing process parameters out of range and acting immediately. Risks during detector filling or operation--such as detector pressure exceeding the setpoint, abnormal LS exchange flux, or abnormal LS temperature--may damage the detector. The safety interlock logic follows two basic principles: independence and fail-safety, which are aligned with the guidance of functional safety standards such as IEC 61508 \cite{IEC61508} and IEC 61511 \cite{IEC61511}.

For the independence principle, the safety logic operates independently of the normal control logic to avoid common-cause failures. For example, the detector calibration house pressure monitoring uses independent sensors in the safety interlock system, ensuring accurate monitoring and triggering of safety actions even if normal control sensors fail. The fail-safety principle means the system will start the safe action by default in case of failure. Critical equipment such as pumps and valves shut down automatically in case of CD pressure imbalance.

Furthermore, the selection of specific equipment also incorporates inherent safety interlock features, thereby reinforcing the overall system's reliability. A prime example is the use of ABB ACS580 \cite{ABB_Manual} variable frequency drives for the electromagnetic pumps. These drives provide a critical, equipment-level layer of protection by continuously monitoring motor current and speed. Should any parameter exceed its predefined safe operating limit, the drive will immediately initiate a safe stop, independently of the higher-level control system. This built-in functionality embodies the fail-safety principle directly at the actuator level and complements the system-wide interlock architecture.

\item Split-range and selective control

Split-range control plays an important role in flow control of the water filling process; a single controller output drives multiple actuators (valve group composed of two regulating valves) to work coherently and precisely. In selective control scenarios, the system implements priority-based logic that dynamically selects control commands from multiple controllers, where critical parameters (e.g., pressure and liquid level) are assigned higher precedence over secondary variables to ensure operational safety.

\end{itemize}

\section{Test and performance}

In alignment with different operational scenarios, multiple simulated condition experiments were conducted to replicate the real FOC working conditions. 

\subsection{Test with pure water system}

Before pure water filling operations, several functional tests were performed in the FOC and water systems using the test pipeline (Fig. \ref{fig:foc_pid}). The incoming water was divided into two streams \cite{mixing}: one feeds directly to the WP, while the other undergoes ultrafiltration for CD use. The water flow is routed through control valves and test pipelines to the backwater, with vortex meters providing PID feedback for valve adjustment.

\begin{figure}[ht]
    \centering
    \begin{subcaptionbox}{WP filling automatic control\label{fig:WP_flux}}
        {\includegraphics[width=0.45\linewidth]{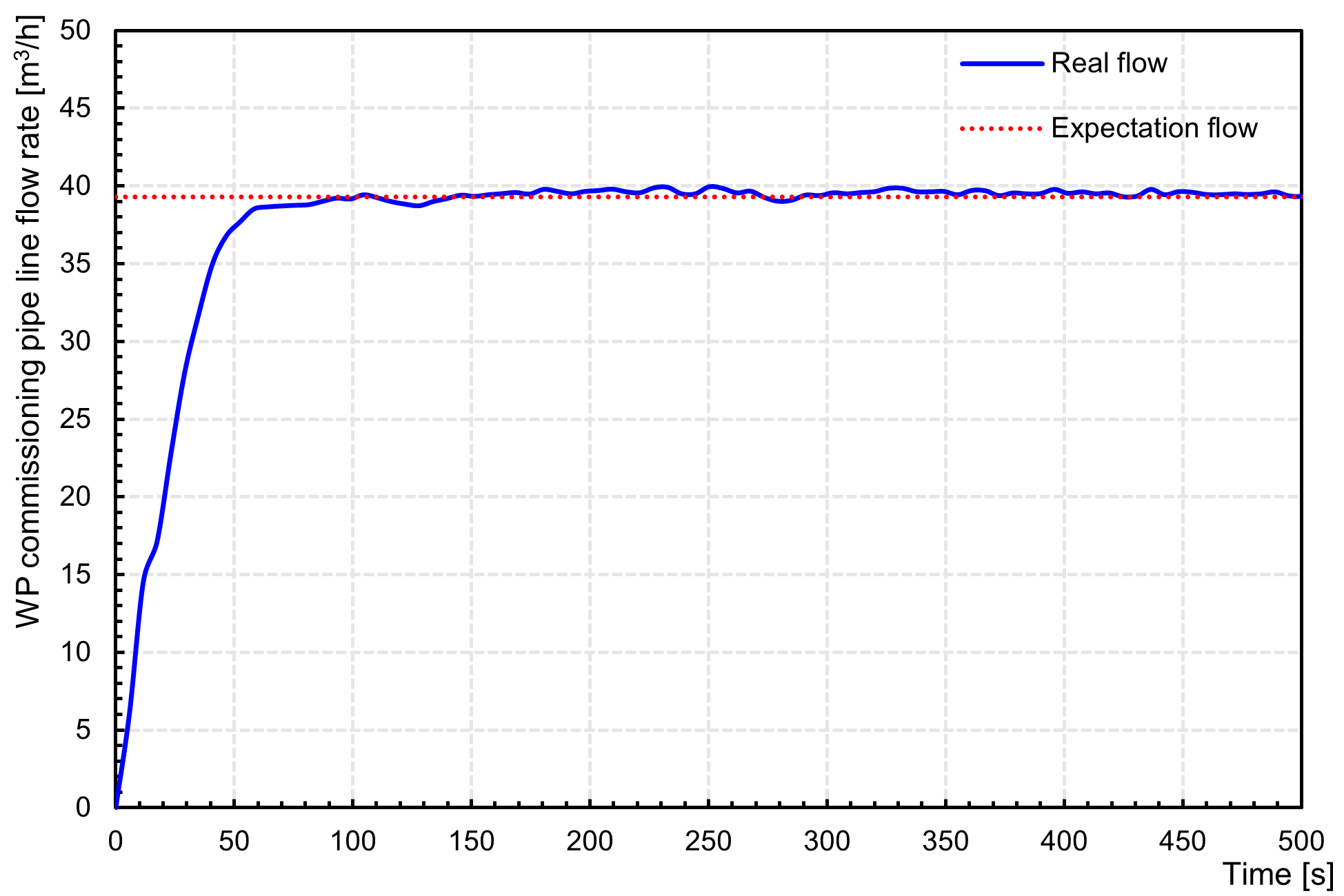}}
    \end{subcaptionbox}
    \begin{subcaptionbox}{CD filling automatic control\label{fig:CD_flux}}
        {\includegraphics[width=0.45\linewidth]{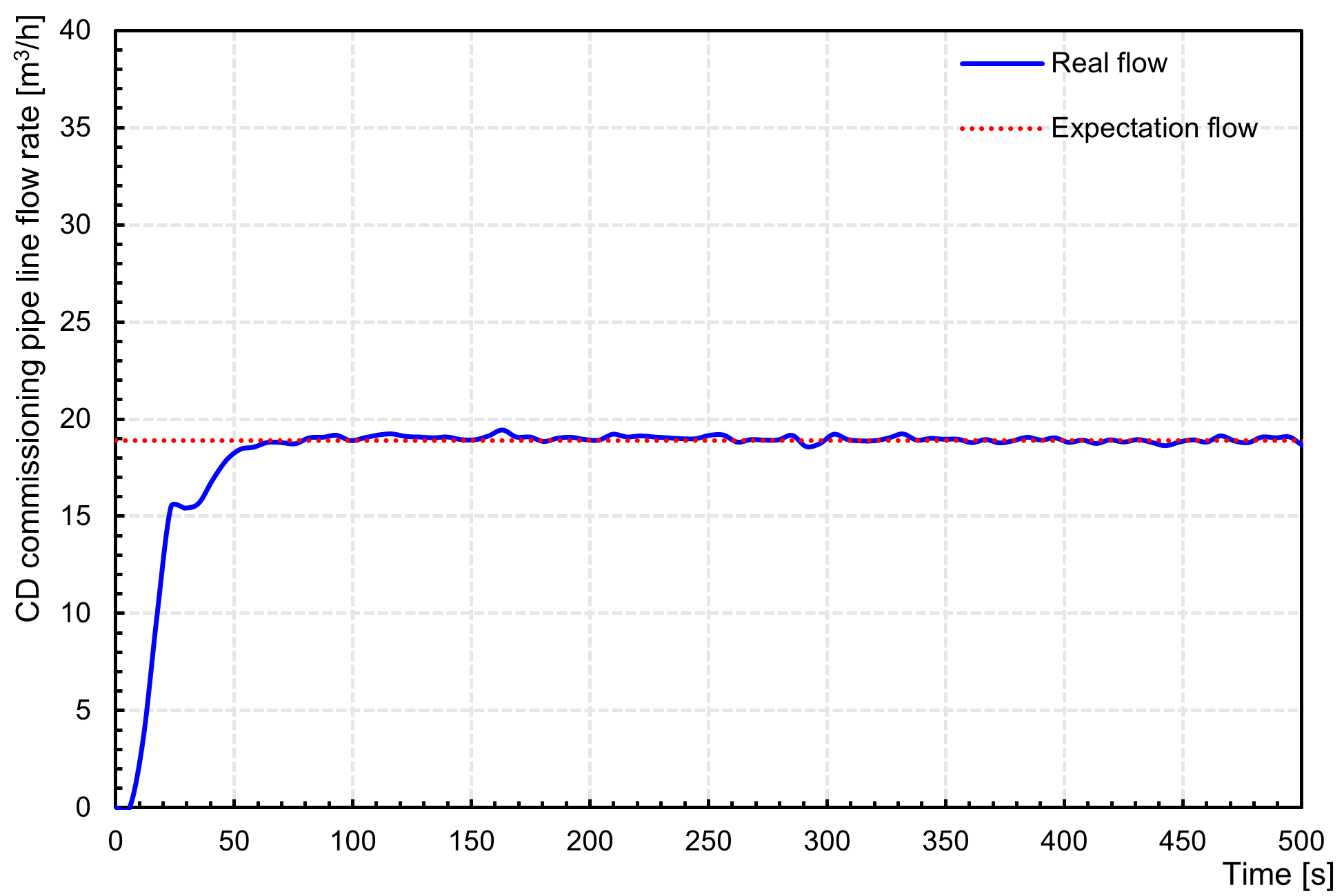}}
    \end{subcaptionbox}
    \caption{WP and CD filling automatic control at test stage}
    \label{fig:PW_flux_comm}
\end{figure}

The test results (Fig. \ref{fig:PW_flux_comm}) demonstrate the rapid response and precise flow control of the control system, with the WP (\(39.3\ \text{m}^3/\text{h}\)) and CD (\(18.9\ \text{m}^3/\text{h}\)) pipelines reached the target flows in 2 minutes and maintained stability within 0.5\% uncertainty during the test period, met all design specifications.

\subsection{Test with LS purification system}

The joint test between the FOC system and the LS purification system successfully verified two operational modes: (1) LS transfer during LS filling (with OSIRIS radioactivity monitoring) and (2) continuous purification recirculation. The test confirmed reliable LS transfer, stable pump performance, and proper execution of sequential control logic and safety interlocks between systems, while maintaining all purity requirements throughout the process.

\begin{figure}[ht]
 \centering
 \makebox[\textwidth][c]{\includegraphics[width=0.65\textwidth]{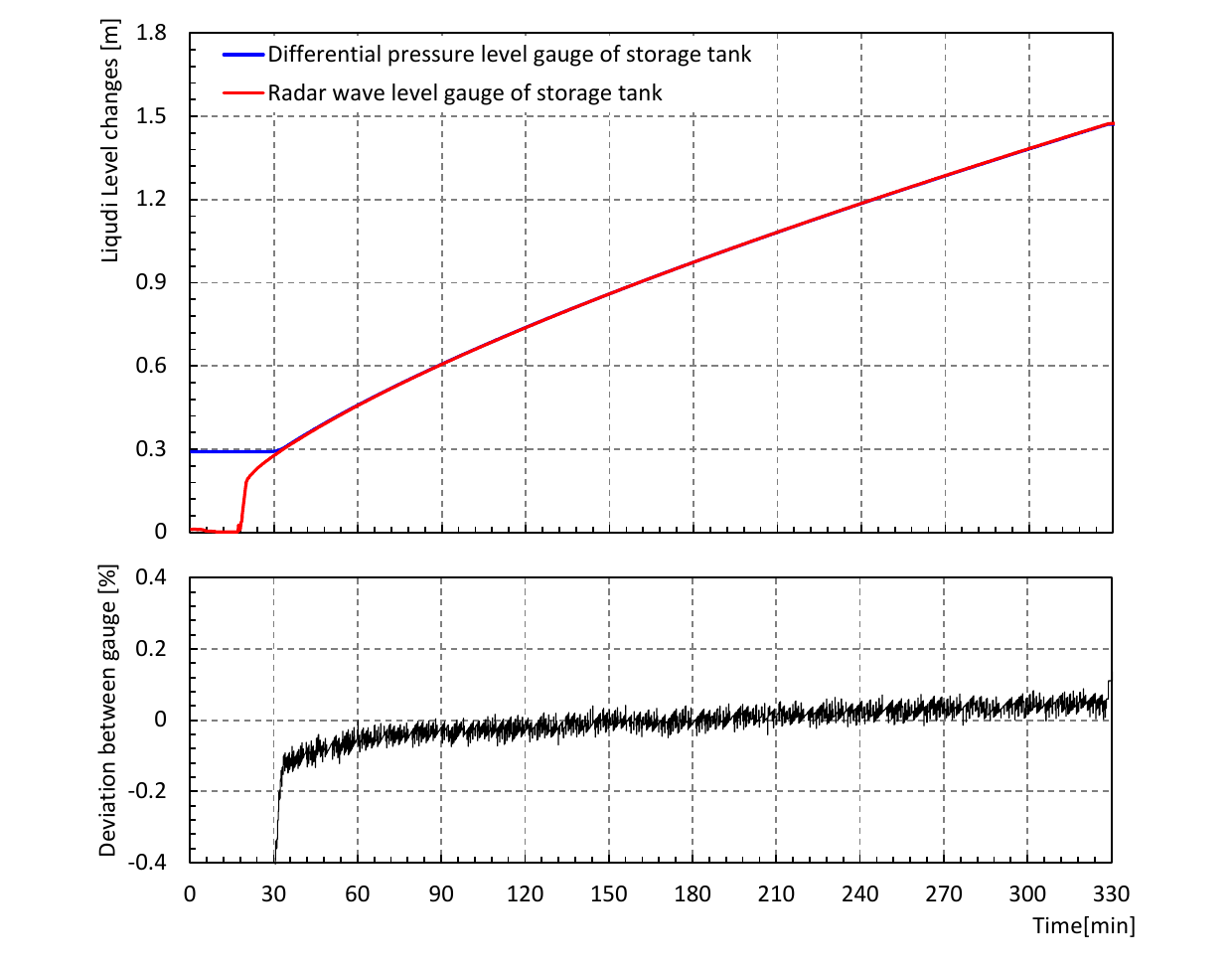}}%
 \caption{Liquid level changes during the test with LS purification system}
 \label{fig:Comm_with_LS}
\end{figure}

During test, LS was transferred from the purification system to FOC tanks at varying flow rates, with automatic pump control triggered by tank level thresholds. Figure \ref{fig:Comm_with_LS} shows the changes in liquid level at \(4.5\ \text{m}^3/\text{h}\) during the transfer of LS between tanks. Although the comparison between differential pressure and radar level gauges confirms measurement accuracy (deviation < 0.2\%), some limitations exist: (1) the differential pressure gauge exhibits a 0-0.3 m dead zone due to installation constraints; (2) The radar gauge demonstrates both a 0.2 m dead zone (for FOC horizontal tanks) and measurable response lag.

\subsection{Internal test of FOC system}

Overflowing test evaluated two key operational modes: (1) overflowing testing through LS transfer from storage to overflow tanks, verifying automatic replenishment when CD/overflow tank levels dropped below lower thresholds, and (2) overfill protection validation, where overflow valves automatically redirected excess LS back to storage after exceeding upper level limits. The tests also confirmed tank connectivity, selective control algorithms, and safety interlocks.

LS filling test utilized the overflow tank and temporary pipelines to validate automatic control performance. LS was pumped from the storage tank through a controlling valve, Coriolis flowmeter, and filter, with PID feedback maintaining the target flow rate ($1500$ L/h) with $0.5\%$ accuracy after a 3-minute stabilization period. This test not only verifies the accuracy matching of the flowmeter, but also visually demonstrates the dynamic response capability of the PID algorithm.


The performance of the control system across all tested scenarios is quantitatively summarized in Table~\ref{tab:performance_summary}. The results demonstrate the system's capability to meet or exceed the design specifications outlined in Section 2.1. Key achievements include the rapid stabilization of flow rates within 1–3 minutes and the maintenance of control accuracy within 0.5\% under both water and LS conditions. The LS transfer test further verified the high precision of the sensor suite, with level gauge deviations remaining below 0.2\% of the full scale. These comprehensive tests confirm the system's robustness, safety, and precision, establishing its readiness for the JUNO detector filling and long-term operation.

\begin{table}[htbp]
\centering
\caption{Summary of key performance metrics from tests}
\label{tab:performance_summary}
\begin{tabular}{lcl}
\toprule
\textbf{Test Scenario} & \textbf{Parameter} & \textbf{Performance Summary} \\
\midrule
Pure Water Filling (WP) & Flow Rate & 
  \begin{tabular}[c]{@{}l@{}}
  Setpoint: $39.3~\mathrm{m^3/h}$, Measured: $39.3 \pm 0.2~\mathrm{m^3/h}$ \\
  Deviation: $\leq 0.5\%$, Stabilized in $1~\mathrm{min}$
  \end{tabular} \\
\addlinespace
Pure Water Filling (CD) & Flow Rate &
  \begin{tabular}[c]{@{}l@{}}
  Setpoint: $18.9~\mathrm{m^3/h}$, Measured: $18.9 \pm 0.2~\mathrm{m^3/h}$ \\
  Deviation: $\leq 0.5\%$, Stabilized in $1~\mathrm{min}$
  \end{tabular} \\
\addlinespace
LS Transfer & Liquid Level &
  \begin{tabular}[c]{@{}l@{}}
  Setpoint: $1500~\mathrm{mm}$, Measured: $1500 \pm 3~\mathrm{mm}$ \\
  Deviation: $< 0.2\%$ FS
  \end{tabular} \\
\addlinespace
LS Filling & Flow Rate &
  \begin{tabular}[c]{@{}l@{}}
  Setpoint: $1500~\mathrm{L/h}$, Measured: $1500 \pm 10~\mathrm{L/h}$ \\
  Deviation: $\leq 0.5\%$, Stabilized in $3~\mathrm{min}$
  \end{tabular} \\
\bottomrule
\end{tabular}
\end{table}

\section{Conclusion}

The FOC system serves as the central platform for LS filling, overflow, and circulation in the JUNO experiment. Integrating a high-stability PLC architecture, precision sensors (accuracy < 0.2\% FS), and advanced control strategies---including PID regulation, sequential logic, and safety interlocks---the system achieves closed-loop control of critical parameters with high reliability. Commissioning tests demonstrate rapid flow stabilization within 3 minutes and a control accuracy of 0.5\% during water/LS filling and transfer operations. Redundancy measures such as modular PLCs, one-operational-one-standby pumps, and software safeguards (e.g., dual HMIs and multi-level alarms) ensure system-level fault tolerance and anti-interference capability.

These accomplishments establish a solid foundation for the safe and efficient filling of the 20-kiloton central detector, directly enabling JUNO's primary scientific missions, including the determination of neutrino mass ordering and high-precision oscillation measurements. The system's ability to maintain long-term LS stability and purity is essential throughout the experiment's projected 20-year lifespan.

Looking ahead, upgrade paths are under consideration to further enhance system performance, such as the implementation of model predictive control (MPC) for process optimization and machine learning--based diagnostic tools for predictive maintenance. Moreover, the FOC system provides a ready-made infrastructure for future physics programs at JUNO, such as neutrinoless double-beta decay searches, which demand extreme LS purity and stability supported by online circulation.

Beyond JUNO, the design principles presented here---including the robust integration of industrial PLCs with experimental control frameworks, hierarchical safety interlocks, and large-volume liquid handling solutions---offer valuable insights for next-generation large-scale detectors, such as Hyper-Kamiokande and DUNE, which face similar liquid-target control challenges.


\acknowledgments

This work is supported by NO.2023YFA1606100, National Key R\&D Program of China.




\bibliography{biblio.bib}
\end{document}